\documentclass[
aps,
prd,
showpacs,amsmath,amssymb]{revtex4}
\usepackage{epsf}
\usepackage{graphicx}  
\usepackage{dcolumn}   
\usepackage{bm}        
\usepackage{color}
\usepackage{hyperref}
\usepackage{epstopdf}
\hypersetup
{
	colorlinks,%
	citecolor=blue,%
	linkcolor=magenta,%
	urlcolor=blue,%
}

\newcommand{\be}{\begin{equation}}
\newcommand{\en}{\end{equation}}
\newcommand{\bea}{\begin{eqnarray}}
\newcommand{\ena}{\end{eqnarray}}
\newcommand{\sch}{Schwarzschild}

\begin{document}
	
	\title{Spontaneous genesis of naked singularities through quantum-gravitational processes: conclusive evidence for violation of cosmic censorship}
	\author{Yang Huang $^1$}\email{sps\_huangy@ujn.edu.cn}
	\author{Hongsheng Zhang $^{1}$}\email{sps\_zhanghs@ujn.edu.cn}
	\affiliation{
		$^1$ School of Physics and Technology, University of Jinan, 336, West Road of Nan Xinzhuang, Jinan 250022, Shandong, China}
	
		\begin{abstract}
	  Cosmic censorship conjecture takes a pivotal status in general relativity. We demonstrate that quantum effects, Hawking effect together with Shwinger effect inevitably  lead to violation of cosmic censorship. We find that naked singularity spontaneously appears in late time evolution of an isolated large dilatonic black hole. The critical discovery is that the power of Hawking radiation converges to a finite value for an extreme dilatonic black hole, which directly exposes the singularity in finite time. The spectrum of Hawking radiation of extreme dilatonic black holes becomes a truncated shrink Planck distribution.  We analyze the underlying physics of the spectrum of Hawking radiation, which roots in extraordinarily wide potential around the horizon. We study the discharge mechanism of a dilatonic black hole through Schwinger effect. Amazingly, the Schwinger pair production naturally ceases for an extreme dilatonic black hole with mass larger than $1.7\times 10^5$ solar masses. Furthermore, we show that evaporation of charged particle because of the Schwinger effect do not save the cosmic censorship for black holes heavier than $1.784\times 10^7 $ solar masses in a significant region of initial charge parameter space. For the first time, we demonstrate that the naked singularities are {\rm spontaneously} formed driven by quantum effects, such that we learn quantum gravity directly from the information shedded from the singularity. 

	\end{abstract}
	
	\maketitle
	
	\section{Introduction}
	The ability to predict is a fundamental characteristic of modern science, particularly physics. Strikingly, Penrose and collaborators proved that singularities inevitably emerge under certain reasonable physical conditions, and whether such a singularity is enveloped by an event horizon remains undetermined \cite{Penrose:1964wq,Hawking_Ellis_1973}. If a naked singularity were to manifest in our universe, it is a disaster for the predictive power of the whole physics since it unpredictably emits particles and information in arbitrary way. To preserve the predictive power of physics in our world, Penrose postulated the cosmic censorship hypothesis to confine the singularity within an event horizon \cite{Penrose:1969pc}. The cosmic censorship conjecture is one of the most important unsolved problems, may be the most one in classical gravity.  
        
    To rigorously prove the cosmic censorship under general conditions is extremely difficult, as is to falsify it. Both of the two fronts of this conjecture has been investigated extensively and deeply, but non general theorem to guarantee this conjecture or an explicit example to violate it which is totally inflexible \cite{Landsman:2021mjt,Wald:1997wa,VandeMoortel:2025ngd}. We have several implications that cosmic censorship may be obeyed in the Universe. A famous study shows that one cannot make a near extreme Kerr-Newman black hole be oversaturate through firing  charged particles to the hole \cite{Wald:1974hkz}. A similar result is derived in dilatonic gravity \cite{Jiang:2019ige}.

     In fundamental level, the matter should be described by wave equation rather than geodesics. Interestingly, collapse of matter wave may lead to naked singularity \cite{Choptuik:1992jv,Guo:2018yyt,Guo:2020ked}. Christodoulou demonstrates that the set of initial conditions which lead to naked singularity is a zero measure set in the total set of whole initial conditions \cite{Christodoulou:1991yfa,Christodoulou:1994hg}. In this sense, naked singularities do not appear for stochastic initial conditions, which saves cosmic censorship.   It seems that the nature tries to hide singularity from our sight. This censorship circumvents the embarrassment from our ignorance of singularity. On the other hand, it seriously impedes our exploration to quantum gravity, which plays critical role only about singularities. The formation of naked singularity deprives the power of predictablity of {\it{classical}}~general relativity, while presents opportunity to probe quantum gravity.  Note that around the horizon of stellar and supermassive black holes, the effects of quantum gravity is, in fact, negligible.
     
     We yet have no explicit evidence for violation of cosmic censorship in frame of classical physics. To directly investigate quantum gravity, it is quite sensible to explore the possibility of peel off or destroy the horizon around classical singularity by quantum effects. 
    The most significant quantum effect for black hole is its Hawking radiation. At first sight, Reissner-Nordstr\"{o}m (RN) black holes appear to potentially expose the singularity through Hawking radiation, since it loses mass of through radiating massless particles including photon, graviton without charge \cite{Hawking:1974rv,Hawking:1975vcx,Page:1976df}. However when a RN black hole approaches extremal state, its temperature approaches zero. And thus its Hawking evaporations cease when an RN black hole becomes extremal.
     
     On the other hand, an isolated RN black hole also presents another significant quantum effect that must be taken into account: the Schwinger effect, which leads to charge loss of the hole. The evolution of RN black holes becomes rather non-trivial when one considers Hawking and Schwinger effects together \cite{Hiscock:1990ex}. For a sufficiently large hole ($M>10^8M_\odot$), 
     the whole evolution of holes with different size is controlled by an attractor, during which the RN black hole approaches near extreme state by never be oversaturate.  Thus the cosmic censorship is saved for an isolated RN black hole.

	One may conjecture that  general charged black holes share similar properties of RN holes, including temperature, entropy, and evolution controlled by Hawking radiation and Schwinger effect. 
    However, the thermodynamics of dilatonic black hole \cite{Garfinkle:1990qj,Gibbons:1987ps} (GMGHS) seems different \cite{Preskill:1991tb}. When the parameter $a<1$ in the dilatonic gravity the evolution of a dilatonic black hole controlled by Hawking radiation is fairly similar to a RN black hole. The temperature goes to zero when a dilatonic black hole with $a<1$ becomes extremal, and thus naked singularity does not appear. A dilatonic black hole with $a>1$ will explode suddenly when it becomes extremal \cite{Koga:1995bs}. The explosion also evades naked singularity in a unusual way.  
    
    The case with $a=1$, which is directly reduced from string theory, is quite dramatic. It is crucial for the whole theoretical frame, but also the most challenging case, or in words in \cite{Holzhey:1991bx}, ``is enigmatic". We decode this enigma and find a striking result, which implies that inevitable violation of cosmic censorship at late time evolution of an isolated dilatonic black hole with $a=1$.   For a near extreme dilatonic black hole, the total power of Hawking radiation of a black hole approaches a finite value rather than zero. Therefore, Hawking radiations spontaneously may lead to naked singularity, which directly ruins the cosmic censorship.  In this article we will show that the real situation is more stunning: the power of Hawking radiation of extreme dilatonic black hole with $a=1$ approaches a finite value while the charge loss because of Schwinger effects ceases, which directly leads to a naked singularity.

     This article is organized as follows. In the next section, we introduce the dilatonic black hole, stressing the fascinating thermodynamic property especially the extreme case and analyzing the property of effective potential of a scalar wave in the dilatonic background. In section III, we conduct a thorough study on Hawking radiation of dilatonic black hole, especially the extreme case. In section IV, we investigate the charge emission process of dilatonic black hole via Schwinger mechanism, and show that such a process ceases for a large extreme hole. We conclude this article and present some related discussions in section V.

  \section{GM black hole}	
In Einstein frame, the action of dilatonic gravity with an electromagnetic field reads \cite{Holzhey:1991bx},
   \be
   S=\frac{1}{16\pi G}\int d^4x \sqrt{-g}\left(R-2\partial_\mu \phi \partial^\mu \phi -e^{-2a\phi}{\cal F}_{\mu\nu}{\cal F}^{\mu\nu}\right),
   \en
  where $\phi$ denotes the dilaton field, $\cal F$ labels the electromagnetic field, and $a\in [0,~\infty)$ is parameter in the theory.

 The spherical spacetime in dilatonic gravity is described by the metric,
	\begin{equation}\label{Eq: metric}
	ds^2=-Fdt^2+F^{-1}dr^2+r^2K\left(d\vartheta^2+\sin^2\vartheta d\varphi^2\right),
	   \end{equation}
	with
	\begin{equation}
	F(r)=\left(1-\frac{r_+}{r}\right)\left(1-\frac{r_-}{r}\right)^{\frac{1-a^2}{1+a^2}},~~~~K(r)=\left(1-\frac{r_-}{r}\right)^{\frac{2a^2}{1+a^2}},
	\end{equation}
    where $r_-$ and $r_+$ are the radiuses of interior and exterior horizons respectively,
    \be
    r_{\pm}=\frac{(1+a^2)(M\pm \sqrt{M^2-(1-a^2)Q^2})}{1\pm a^2}.
    \en
	Here $M$ and $Q$ are the mass and electric charge of the black hole, respectively.
	The corresponding dilaton and electric potential read,
	\be
	e^{\phi}=e^{\phi_0}\left(1-\frac{r_-}{r}\right)^{\frac{a}{1+a^2}},
	\en
	and,
	\be
	A_d=\frac{Q}{r}dt,
	\label{epotential}
    \en
	respectively.
	$\phi_0$ is the value of the dilaton field $\phi$ at spacelike infinity. The value of $\phi_0$ implies the asymptotic behavior of the manifold. In this article we consider an asymptotically flat manifold, which requires $\phi_0=0$.
	
	The temperature derived from surface gravity of dilatonic black hole is written as,
	\be
	T=\frac{1}{4\pi r_+} \left(1-\frac{r_-}{r_+}\right)^{\frac{1-a^2}{1+a^2}}.
	\label{temp}
	\en
	The Hawking radiation of dilatonic black hole for both the cases with $a>1$ and $a<1$ has been studied to some extent \cite{Koga:1995bs}. However, extra complexity is involved in  the case with $a=1$, 
  especially for an extreme hole.  Mathematically, $0^0$ is not well-defined. Its value depends on the paths of repeated limits. If one first takes an extreme black hole and after that lets $a\to 
 1$, then $T=0$.
    If one first takes $a=0$ and after that lets $Q\to Q_{\rm max}$, then $T=\frac{1}{4\pi r_+}$. Explicitly, we have
    \be
    T_{\rm extreme~path1}=\lim_{a\to 1}\left(\lim_{~~Q\to Q_{\rm max}}T\right)=0,
    \label{T1}
    \en
    while,
    \be
    T_{\rm extreme~path2}=\lim_{~~Q\to Q_{\rm max}}\left(\lim_{a\to 1}T\right)=\frac{1}{4\pi r_+}.
    \label{T2}
    \en
    Surprisingly, this remarkable property has never been mentioned in extensive investigations of thermodynamics of dilatonic gravity. Fundamentally, we decipher the real temperature  only after we show the corresponding distribution of Hawking radiations.  Amazingly, we find that both of the two paths partly tell the truth. For  radiations with $\omega<\omega_0$, the extreme dilatonic black hole behaves like a black body with $T=0$, while for radiations with $\omega \geq\omega_0$, it seems like a grey body with  $T=\frac{1}{4\pi r_+}$. One will see that the resulted distribution is a truncated shrink Planck distribution.  We demonstrate this point in the following text in detail. 
    
    In the following discussions we concentrate on the case $a=1$.

	For convenience, we introduce a normalized charge $q=Q/Q_{\mathrm{max}}\in\left[0,1\right]$, and parameterize the charge by $q=1-e^{-\eta}$. The Schwarzschild black hole corresponds to $\eta=0\;(q=0)$, while the extreme dilatonic black hole corresponds to $\eta\rightarrow\infty$ $(q\rightarrow1)$.
	
	Scattering and bound states of a massless scalar field by the dilatonic black hole has been studied in \cite{Huang:2020bdf}. To explore Hawking radiations, we investigate wave equation with different boundary conditions. The Klein-Gordon equation of a massless scalar field reads $\nabla_\mu\nabla^\mu\Psi=0$. This equation admits the following separable solutions of the form \cite{Huang:2020pga},
	\begin{equation}\label{Eq: separate variable}
	\Psi=e^{-i\omega t}\frac{\psi_{\omega l}(r)}{r\sqrt{K(r)}}Y_{lm}(\vartheta,\varphi),
	\end{equation}
	where the radial function $\psi$ obeys the radial equation,
	\begin{equation}\label{Eq: the radial eq}
	\frac{d^2\psi_{\omega l}}{dx^2}+\left[\omega^2-V_l(r)\right]\psi_{\omega l}=0,
	\end{equation}
	with $x=\int dr/F$ the tortoise coordinate, and the effective potential given by,
	\begin{equation}\label{Eq: effective potential}
	\begin{aligned}
	V_l(r)=&\frac{F(r)}{K(r)}\left[\frac{F'(r)}{r}+\frac{l(l+1)}{r^2}\right]\\&-\frac{2M^2q^2}{r^4}\frac{F(r)}{K(r)^2}\left[1+\frac{q^2}{2}\left(1-\frac{6M}{r}\right)\right].
	\end{aligned}
	\end{equation}
	
\begin{figure}
		\centering	
		\includegraphics[width=0.45\textwidth,height=0.35\textwidth]{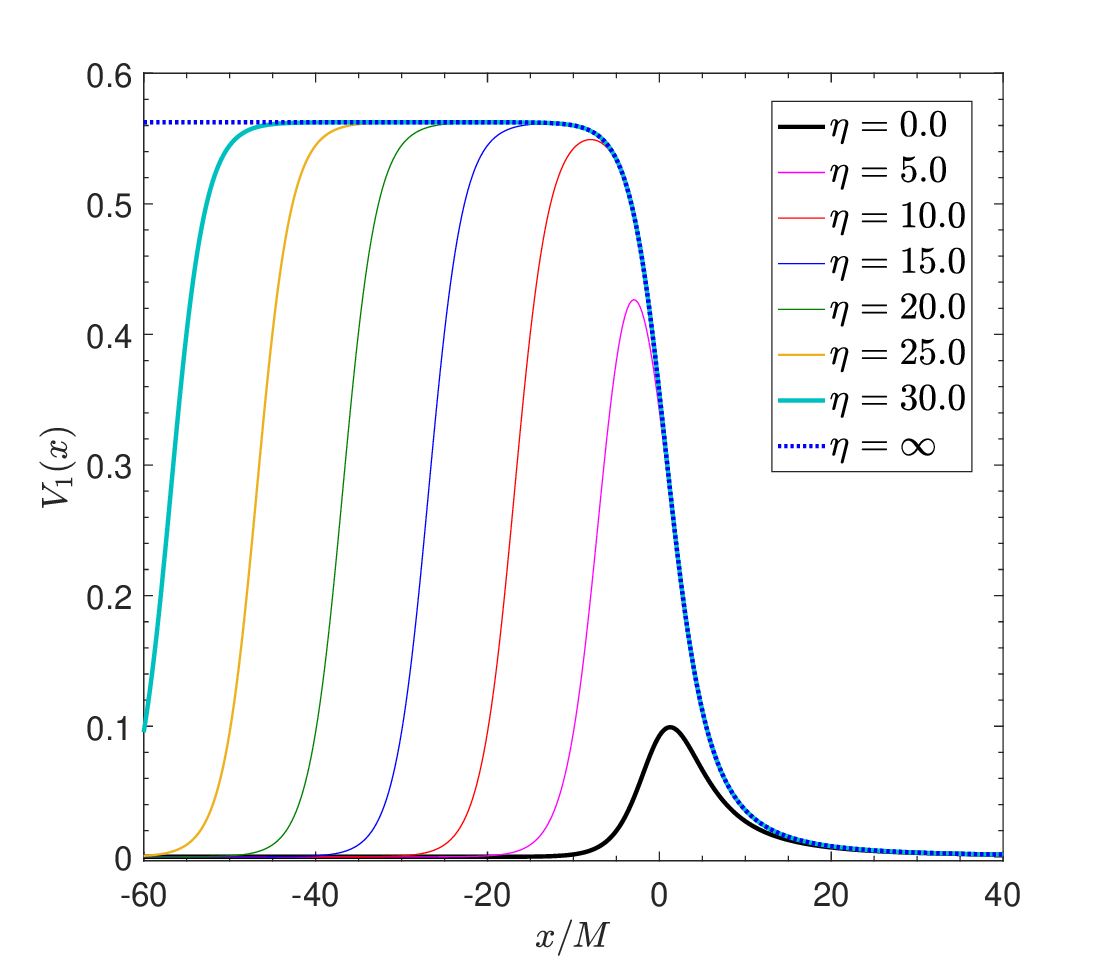}		
		\caption{Comparison of the effective potential, with $l=1$, for different $\eta$.}
		\label{Fig: potential}
	\end{figure}

	In Fig. \ref{Fig: potential}, we compare $V_{l=1}(x)$ for different values of $\eta$.
	We see that the width of the potential barrier increases monotonously with the increase of $\eta$.
	For $\eta>10$, $V_l(x)$ is similar to a rectangular potential barrier.
	Furthermore, in the extreme limit $\eta\rightarrow\infty$, the width of the potential barrier increases without bound, whereas the height tends to $(2l+1)^2/16M^2$ \cite{Huang:2020pga}.
	
 \section{Hawking radiation of dilatonic black hole}
 
  In this section, we first conduct a general study of Hawking radiation of dilatonic black hole, concentrating on the behaviour of the spectrum when the hole becomes extremal. Then, we investigate the Hawking radiation from an exact extreme hole, and prove that the spectrum of Hawking radiation of the exact extreme hole is the limit of the power of a non-extreme one as $q\to 1$.
 
To investigate the Hawking radiation detected by observers at infinity, we first study the absorption of scalar field by the dilatonic black hole. Then, the asymptotic solutions of Eq.(\ref{Eq: the radial eq}) at the horizon and infinity are given by,
	\begin{equation}\label{Eq: absorption}
	\psi_{\omega l}\sim
	\left\{
	\begin{aligned}	
	&\mathcal{T}_{\omega l}e^{-i\omega x},\;\;\;\;&\text{for}\;\;r\rightarrow r_+,\\
	&e^{-i\omega x}+\mathcal{R}_{\omega l}e^{i\omega x},\;\;\;&\text{for}\;\;r\rightarrow\infty,
	\end{aligned}
	\right.
	\end{equation}
	where $\mathcal{T}_{\omega l}$ and $\mathcal{R}_{\omega l}$ are the transmission and reflection coefficients, respectively. They satisfy the conservation law
	\begin{equation}
	|\mathcal{R}_{\omega l}|^2+|\mathcal{T}_{\omega l}|^2=1.
	\label{conser}
	\end{equation}
	where $|\mathcal{T}_{\omega l}|^2$ and $|\mathcal{R}_{\omega l}|^2$ denote transmission and reflection probabilities, respectively. We obtain $|\mathcal{T}_{\omega l}|^2$ by numerically integrating the radial equation (\ref{Eq: the radial eq}) with boundary conditions given in Eq.(\ref{Eq: absorption}).  Then, the total absorption cross section of the scalar field is given by \cite{PhysRevD.14.3251},
	\begin{equation}\label{cross section}
	\sigma=\frac{\pi}{\omega^2}\sum_{l=0}^{\infty}(2l+1)\left(1-|\mathcal{R}_{\omega l}|^2\right).
	\end{equation}
	The above equation holds true for an extreme hole, see a detailed discussions in Appendix A. This is a critical observation which encodes the property of Hawking radiation of extreme dilatonic black holes.
	
\begin{figure*}
		\centering	
		\includegraphics[width=0.45\textwidth,height=0.35\textwidth]{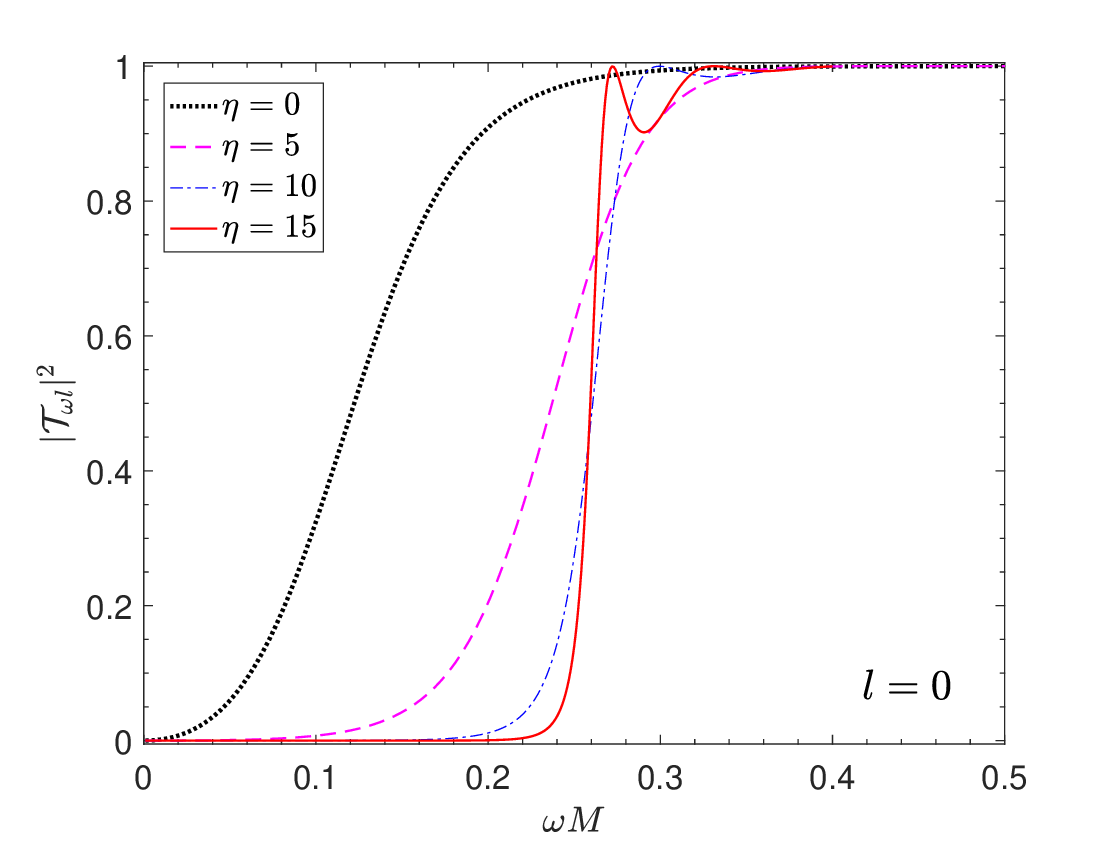}	
		\includegraphics[width=0.45\textwidth,height=0.35\textwidth]{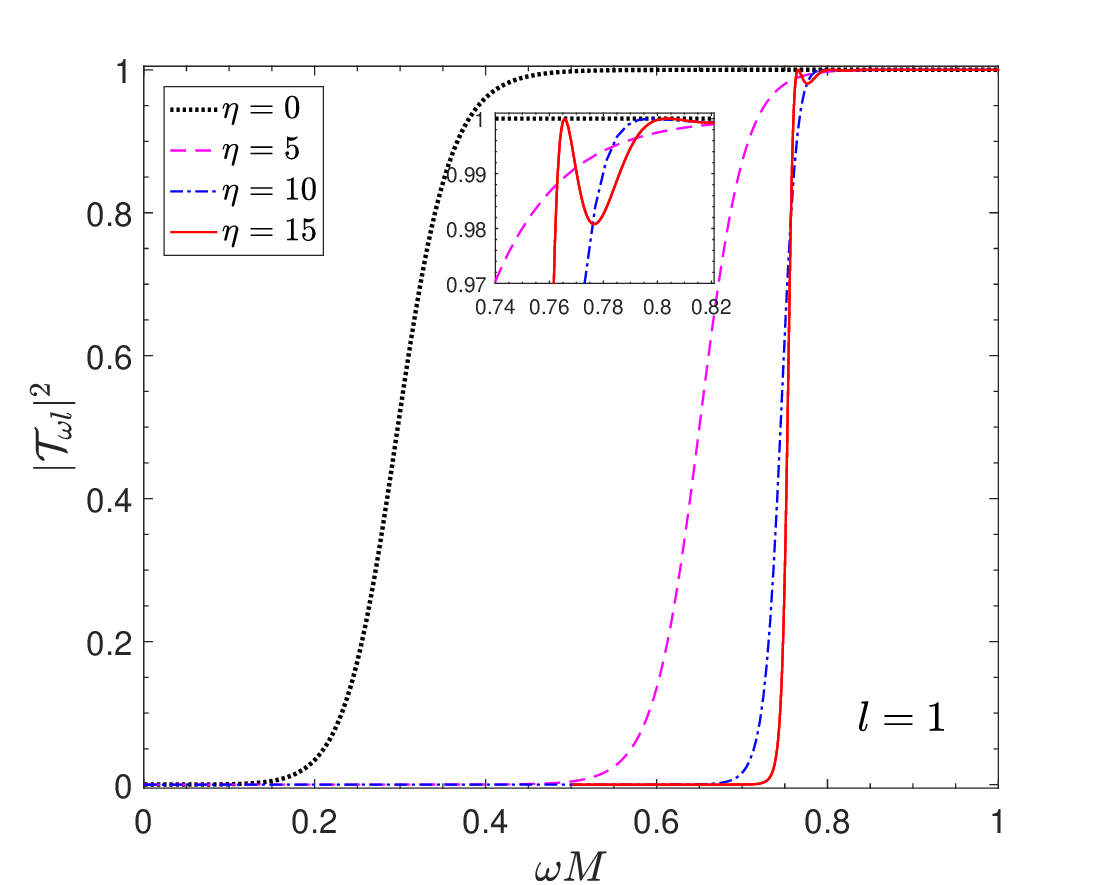}	
		\caption{Transmission probability of the scalar field in the dilatonic black hole spacetime, for $l=0$ (left panel) and $l=1$ (right panel).}
		\label{Fig: transmission 0}
	\end{figure*}

	Figure \ref{Fig: transmission 0} shows the dependence of transmission probability on $\omega$, for different values of $\eta$. We see that $|\mathcal{T}_{\omega l}|^2$ vanishes for $\omega M\rightarrow0$, whereas tends to $1$ in the high frequency limit. Another interesting observation is that for $\omega M<(2l+1)/4$, $|\mathcal{T}_{\omega l}|^2$ decreases with increase of $\eta$. And in the limit of $\eta\rightarrow\infty$, we have $|\mathcal{T}_{\omega l}|^2\rightarrow0$, which means that such low frequency waves will be totally reflected to infinity. This result is consistent with the potential analysis in Fig. \ref{Fig: potential}. Finally, we see that when $\omega M>(2l+1)/4$ and $\eta\gg1$, $|\mathcal{T}_{\omega l}|^2$ oscillates with $\omega M$. As we discussed above, the potential behaves like a rectangular barrier, and the oscillation in $|\mathcal{T}_{\omega l}|^2$ is resulting from the resonance transmission of the rectangular barrier.

\begin{figure}
		\centering	
		\includegraphics[width=0.45\textwidth,height=0.35\textwidth]{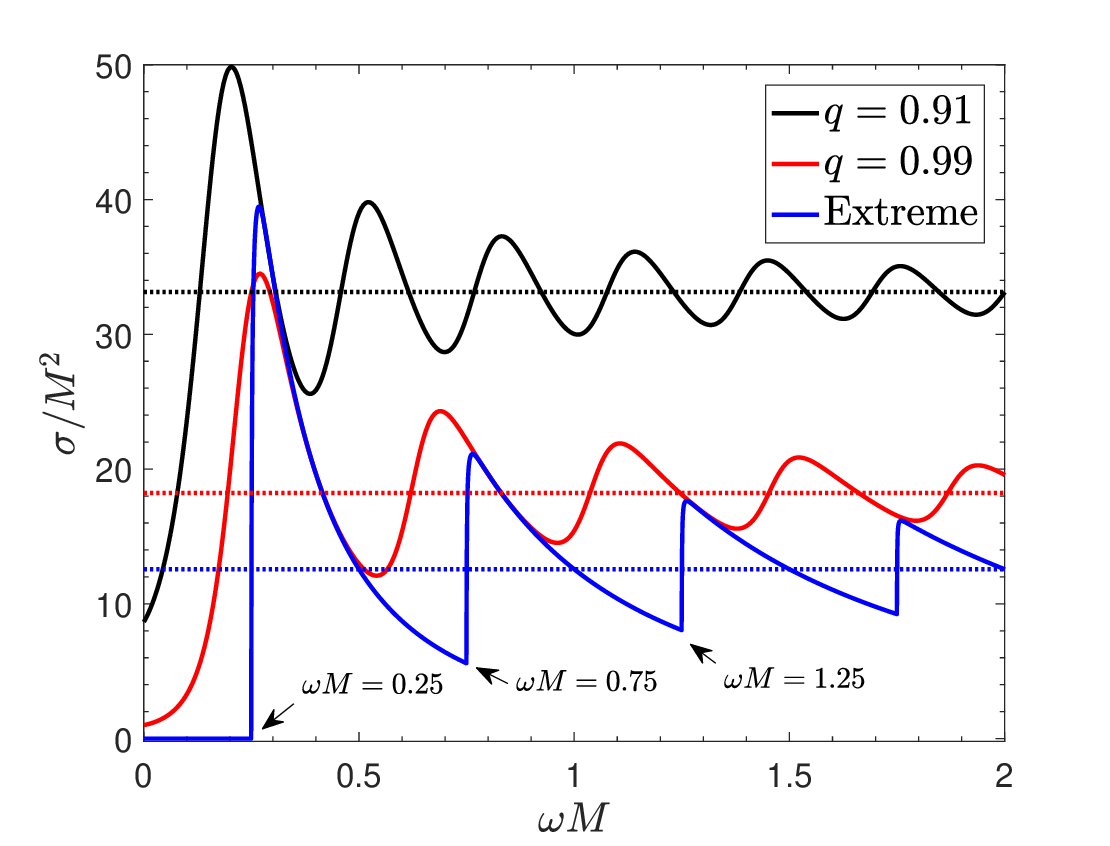}		
		\caption{Total absorption cross sections of massless scalar field, for different values of $q$. The dotted lines in the plot are given by Eq.(\ref{Eq: hf limit}). The position of zigzags of the cross section of the extreme case is commanded by (\ref{zigzagl}).}
		\label{Fig: absorption}
	\end{figure}

	In Fig. \ref{Fig: absorption}, we present typical results of the absorption cross section. We see that for each case, $\sigma$ oscillates around its high frequency limit,
	\begin{equation}\label{Eq: hf limit}
	\sigma^{hf}=\pi b^2_{c}=\pi M^2\left[18-14q^2+\frac{\xi}{2}\left(9-q^2\right)-\frac{\xi^2}{2}\right],
	\end{equation}
	where $\xi=\sqrt{q^4-10q^2+9}$, and the expression of $b_c$ can be found in Eq.(7) in Ref. \cite{Huang:2020bdf}. One may check that for $q=0$, we have $\sigma^{hf}=27\pi M^2$. For $q=1$, one has $\sigma^{hf}=4\pi M^2$. In both cases, $\sigma^{hf}$ reduces to the photon spheres of the holes.
	Let us focus on the extreme case, in which one sees a zigzag pattern in the total absorption cross section. Whenever 
 \be 
 \omega M=(2l+1)/4, ~~l=0,1,2,\cdots,
 \label{zigzagl}
 \en
  there is an abrupt increase in $\sigma$. Particularly, when $\omega M<0.25$, we arrive at $\sigma=0$. This is consistent with the fact that the horizon area of an extremal dilatonic black hole is zero, since a long wave cannot sense the extreme black hole.

    We emphasize that the total absorption section converges (\ref{cross section}) to a finite value when the dilatonic hole goes to be extreme, although the area of horizon vanishes for such a hole. According to general theory of quantum absorption of black holes, the absorption cross section goes to the area of horizon at low energy limit, while goes to the area of photon sphere at high energy limit. Thus it is reasonable that an extreme dilatonic hole has a non-vanishing absorption cross section, but a vanishing absorption cross section when $\omega M \to 0$.
	
	Let us now consider the Hawking radiation of massless scalar particles from dilatonic black holes. We start from non-extreme black holes ($q<1$, or $\eta$ takes a finite value). The principle equation to describe Hawking radiation is the same as (\ref{Eq: the radial eq}), whereas the boundary conditions at the horizon and infinity are different from (\ref{Eq: absorption}). Consider an outgoing wave from the horizon, and then scattered by the potential barrier. This process is described by
	\be\label{hawk}
	\psi_{\omega l}\sim
	\left\{
	\begin{aligned}	
		&\mathcal{B}_{\omega l} e^{i\omega x},\;\;\;\;&\text{for}\;\;r\rightarrow \infty,\\
		&e^{i\omega x}+\mathcal{A}_{\omega l}e^{-i\omega x},\;\;\;&\text{for}\;\;r\rightarrow r_+.
	\end{aligned}
	\right.
	\en
	where $\mathcal{A}_{\omega l}$ and $\mathcal{B}_{\omega l}$ can be viewed as the reflection and transmission coefficients, respectively. One  finds that $|\mathcal{B}_{\omega l}|^2=|\mathcal{T}_{\omega l}|^2$. Routinely, the spectral of Hawking radiation is given by,
	\be
	N_l(\omega)=\frac{\Gamma_{\omega l}}{e^{8\pi M\omega}-1},
	\label{distri}
	\en
	where $\Gamma_{\omega l}=(2l+1)\left(1-|\mathcal{R}_{\omega l}|^2\right)$ is the absorption probability of the scalar field. Then, the mass loss rate of the black hole is given by
	\be
	\frac{dM}{dt}=-\sum_{l=0}^{\infty}\frac{1}{2\pi}\int_{0}^{\infty}d\omega\omega N_l(\omega).
	\label{powerH}
     \en

\begin{figure}
		\centering	
		\includegraphics[width=0.45\textwidth,height=0.35\textwidth]{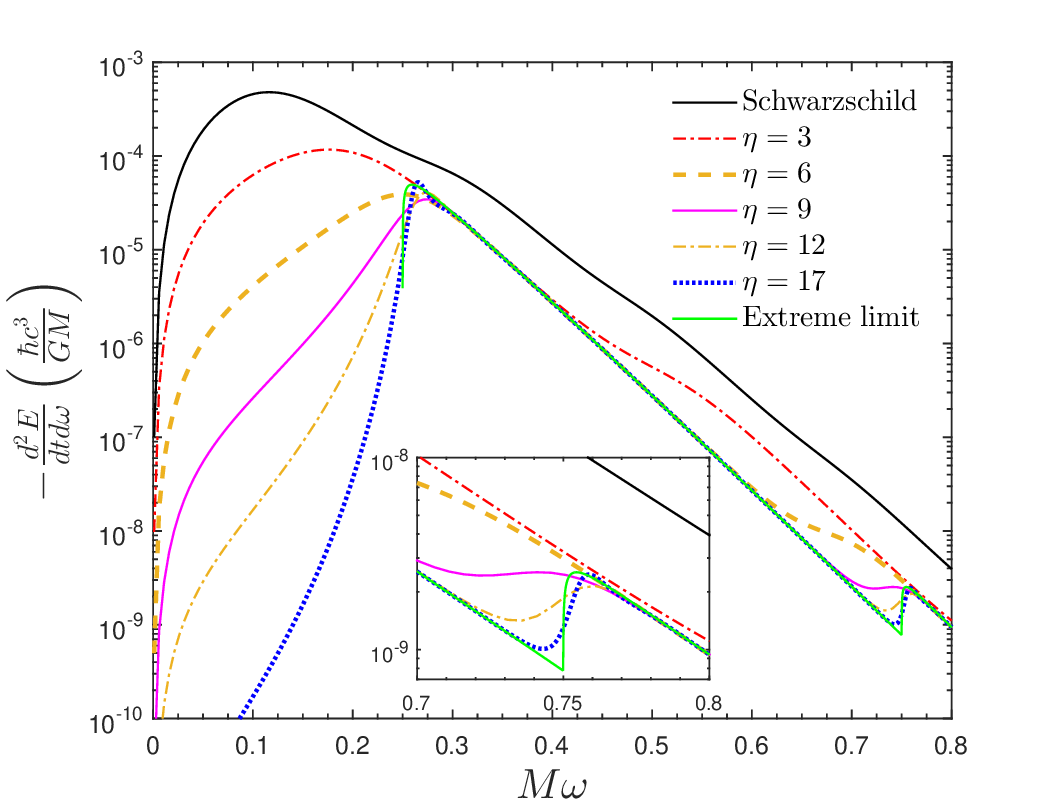}		
		\caption{Power spectra of massless scalar field emitted from dilatonic black holes for different $\eta$. The inset panel displays the behaviour of the spectra around $l=0.75$, where a clear zigzag emerges when the black hole becomes extremal. A similar behaviour appears at $l=0.25$, as the theoretical result implied (\ref{zigzagl}).     }
		\label{Fig: power of HK0}
	\end{figure}

	To display the effects of charge of dilatonic black hole in Hawking radiation, we plot Fig. \ref{Fig: power of HK0}, which shows the numerical results of the power spectra of massless scalar field for different values of $\eta$, in which the case of Schwarzschild corresponds to $\eta=0$. This figure highlights that: (i) the emission rate is suppressed by larger $\eta$; (ii) in the near extreme limit $\eta\gg1$, the spectra have abrupt increases for $\omega M=0.25$ and $0.75$, corresponding to $l=0$ and $l=1$, respectively. In the extreme case, the radiations with $\omega M<0.25$ is completely blocked by the potential around the horizon, and the outgoing waves with higher frequencies are suppressed such a potential. Therefore, both temperatures (\ref{T1}) and (\ref{T2}) partially tell the truth: (\ref{T1}) corresponds to the low frequency limit, while (\ref{T2}) corresponds to high frequency limit.  We name this distribution as truncated shrink Planck distribution.     
	
\begin{figure}
		\centering	
		\includegraphics[width=0.45\textwidth,height=0.35\textwidth]{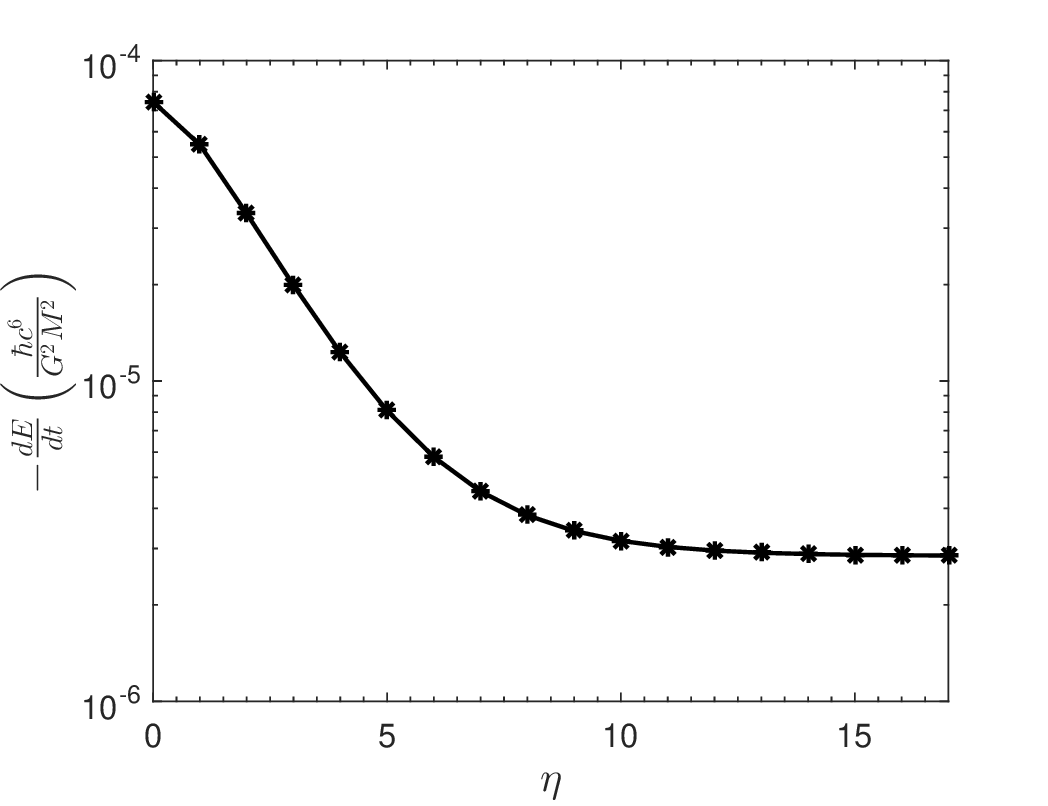}		
		\caption{Total emission power as a function of $\eta$.}
		\label{Fig: power Vs eta}
	\end{figure}

	In Fig. \ref{Fig: power Vs eta}, we present the total emission power as  functions of $\eta$. Again, the total power decreases with increase of $\eta$. The emission rate of the Schwarzschild black hole is about $26$ times that of the dilatonic black hole for $\eta=15$. We find that in the limit of $\eta\gg1$, the emission power tends to $2.85\times10^{-6}\hbar c^6G^{-2}M^{-2}$. The key point is that the total power of Hawking radiation does not vanish for an extreme hole, which is completely from the RN case.  Thus we arrive at the conclusion that for a near extreme dilatonic black hole the naked singularity may {\it spontaneously appears} because of Hawking radiation.
	
 Now we consider the radiation from an exact extreme hole, for which the effective potential becomes $V_l=(2l+1)^2/16M^2$  at the horizon, where, $l=0,1,2,\cdots$. 
 Thus, Eq. (\ref{Eq: the radial eq}) becomes,
	\be
	\frac{d^2\psi_{\omega l}}{dx^2}+\left(\omega^2-V_l\right)\psi_{\omega l}=0.
	\en
	For $\omega<\sqrt{V_l}$, the ``wave" becomes a decaying one, $\psi_{\omega l}\sim e^{-x\sqrt{V_l-\omega^2}}$.
	Only waves whose frequencies $\omega\geq m_l$ have opportunities to propagate to infinity. For this case, the outgoing wave with time $\varphi$ reads,
	\be
	\varphi=e^{-i\omega t}{\psi_{\omega l}(r)}=e^{-i\left(\omega t-k_l x\right)},
	\en
	where $k_l=\sqrt{\omega^2-V_l}$ is a real number for $\omega^2 \geq V_l$.  In {\it all} previous explorations of Hawking radiations of black holes, the ingoing and outgoing particles behave as massless ones regardless of their real masses. On the contrary, all the particles propagating through the horizon behave like massive ones  regardless of their real masses for an extreme dilatonic hole.

 By using arguments similar to the Hawking' original demonstration \cite{Hawking:1975vcx},   one obtains the spectrum of distribution of the radiations,
	\be
	N_l(\omega)=
	\left\{
	\begin{aligned}
		&0,~~&\omega<\sqrt{V_l},\\
		&\frac{\Gamma_{\omega l}}{e^{8\pi M\omega}-1},~~&\omega>\sqrt{V_l},
	\end{aligned}
	\right.
	\label{distri}
  \en

	This is not a familiar distribution. We name it truncated grey distribution, because waves with the frequency less than $\sqrt{V_l}$ are completely locked in the potential barrier. For such waves, an extreme dilatonic black hole behaves frozen star without quantum radiation. This property is also confirmed by our previous studies of bound state of dilatonic black hole, where a wave with some low frequency is stabilized outside the hole by the potential barrier \cite{Huang:2020pga,Huang:2021nii,Huang:2022dfx}. In fact, the existence of true bound states and truncated spectrum are dual effects, both giving prominence of the potential with  infinite width but finite height.

 The total power of the radiation is calculated by (\ref{powerH}), which presents,
  \be
  \frac{dM}{dt}=-2.85\times10^{-6}\frac{\hbar c^6}{G^{2}M^{2}}.
  \label{pHradia}
  \en 
  For convenience of numerical calculation, we recover all the physical constants in the above equation. One sees that the radiation power of a non-extreme dilatonic black hole exactly converges to that of the extreme hole when $q\to 1$.

\begin{figure*}
		\centering
		\includegraphics[width=0.95\textwidth,height=0.25\textwidth]{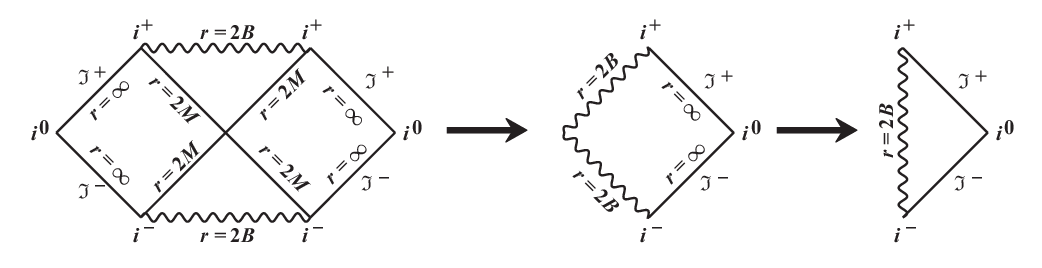}	
		\caption{Penrose diagrams of ordinary, extreme, and over charged dilatonic black holes, from the left panel to right panel. $B=\frac{Q^2}{2M}$ in this figure. Hawking radiation exposes the singularity of a dilatonic black hole.  An ordinary dilatonic black hole spontaneously evolves to be an extreme hole and then a naked singularity because of Hawking radiation.}
		\label{Penrose3}
	\end{figure*}

To illustrate the changes in the global structure of a dilatonic black hole from a non-extreme black hole to an extreme one, then to a naked singularity because of Hawking radiation, we plot Penrose diagrams in Fig. \ref{Penrose3} for this process. After even emission of even a single charge-free Hawking particle, it will becomes a true naked singularity as shown in the right panel in Fig. \ref{Penrose3}.


  \section{considering charge emission}

  In previous discussions, we do not consider Hawking evaporation of charged/chargeless massive particles, nor other possible process yielding discharge of the hole. For a black hole heavier than $10^{17}$g, or its temperature lower than $10^9$K, it in fact only radiate massless particles \cite{Page:1976df}. The emission of massive particles via Hawking effects is safely omitted. Because of there is no charged massless particle, it does not yield charge loss through Hawking radiation for a large black hole. It seems that the charge mass ratio will increase for an isolated massive black hole through Hawking evaporation. 
  
   For massive black holes, a different mechanism to discharge, that is, the Schwinger mechanism \cite{Schwinger:1951nm}, may become important. We investigate this Schwinger effect in the background of dilatonic black holes. If the size of a black hole is much larger than the Compton wavelength of the electron, the pair production can be soundly described by QED in Minkowski space \cite{Gibbons:1975kk}. According to the Schwinger effect the pair production rate of electron-position per unit 4-volume reads,
          
          \be
          {\cal G}=\frac{(eE)^2}{4\pi^3\hbar^2}\left[1+{\cal O}\left(\frac{e^3E}{\mu^2}\right)\right] {\rm exp} \left(\large{-\frac{\pi  \mu^2}{eE\hbar}}\right),
          \label{calG}
          \en  
          where $e,~\mu$ denote the charge and mass of an electron respectively, and $E$ for external electric field strength at that position.
   According to (\ref{epotential}), the electric strength of a dilatonic black hole reads,
   \be
   E=\frac{Q}{r^2}.
   \en
   Thus, the pair production rate reads,
   \be 
             {\cal G}=\frac{(eQ)^2}{4r^4\pi^3\hbar^2}\left[1+{\cal O}\left(\frac{e^3Q}{r^2\mu^2}\right)\right] {\rm exp} \left(\large{-\frac{\pi  r^2\mu^2}{eQ\hbar}}\right).
          \label{calGp}
          \en 
  
  
  
   
   
   By introducing two constants $K_1$ and $K_2$, we have
  \be
  {\cal G}=\frac{K_1}{r^4}\exp (-K_2r^2),
  \en
  where,
  \be
  K_1=\frac{e^2Q^2}{4\pi^3\hbar^2},~~K_2=\frac{\pi \mu^2}{\hbar eQ}.
  \en 
  
  The production rate of electric charge reads,
  \be
   {\cal G}_e=e{\cal G}.
   \en
   
   The total production rate of electric charge in a dilatonic black hole spacetime reads,
   \bea 
   \frac{dQ}{dt}&=&-4\pi \int_{r_+}^{\infty} dr r(r-r_-) {\cal G}_e \nonumber\\
   &=&-4\pi e K_1 \left[\frac{\exp(-K_2r_+^2)}{r_+}-\sqrt{\pi K_2} ~\text{erfc}(\sqrt{K_2}r_+)\right]+4\pi e \frac{K_1 r_-}{2r_+}\left[\frac{\exp(-K_2r_+^2)}{r_+}+K_2r_+Ei(-K_2r_+^2)\right],
   \label{volume} \ena 
   where $r_-=\frac{Q^2}{M}$, erfc labels the the error function, and $Ei$ labels the exponential integral function. The first term is exactly the result of RN black hole \cite{Hiscock:1990ex}, and the second term indicates the surplus effects of the dilaton field. 
   For large black holes, we expand the error function $\text{erfc}$ for large argument,
   \be 
   \text{erfc}(x)=\frac{\exp (-x^2)}{\sqrt{\pi}}\left(\frac{1}{x}-\frac{1}{2x^3}+\frac{3}{4x^5}+...\right),
   \en 
   and the exponential integral function for large argument,
   \be
   Ei(x)=\exp(x)\left(\frac{1}{x}+\frac{1}{x^2}+\frac{2}{x^3}+...\right).
   \en

   We consider black holes heavier than $Q_0$, then the total production rate of electric charge becomes,
   \bea
   \frac{dQ}{dt}
   =-\frac{e^4Q^3}{2\pi^3\hbar \mu^2}\frac{1}{r_+^3} \left(1-\frac{r_-}{r_+}\right)\exp \left(-\frac{r_+^2}{QQ_0}\right), ~~~Q_0=\frac{\hbar e}{\pi \mu^2}=1.7\times 10^5M_{\odot}.
   \label{Qevolve}
   \ena
  
  One sees that the rate charge production is deficiency than the corresponding RN black with the same mass and charge. Physically, the effective volume shrinks since $r$ is replaced by $r-r_-$ in the integral (\ref{volume}). When the black hole becomes extreme, the process of charge production completely ceases since $r_-=r_+$. In previous section, we prove that the power of Hawking radiation is a finite value, see (\ref{pHradia}). One can always bring a dilatonic black hole into a near-extremal state by throwing charge into it \cite{Jiang:2019ige}. No matter how the black hole reaches a nearly extremal state, it will spontaneously expose the singularity through Hawking radiation, as its Schwinger effect completely ceases at this point.
  
  More impressively, for a large black hole ($M>10^8 M_{\odot}$) we prove that a naked singularity will spontaneously emerge  through the interplay of Hawking radiation and the Schwinger effect under a significant range of initial charge parameters of the hole.  The mass loss is separated into two sectors. The first one is the Hawking radiation of massless neutral particles, and the second one is due to the Schwinger effects. The first sector of massless reads,
  \be
  \frac{dM_1}{dt}=-aT^4\sigma_0 \xi {\cal A} (\eta).   
  \en
  Here,
  \be
  a=\frac{\pi^2}{15\hbar^3},
  \en
  \be
  T=\frac{\hbar}{8\pi M},
  \en 
  \be 
  \sigma_0={27}\pi M^2,
  \en 
  $\xi$ is related the kinds and properties of massless particles. The massless particles include photon, graviton, and possible electron neutrino. We are still uncertain whether the electron  neutrino has mass. Regardless of whether electron neutrinos are massive, this parameter is a quantity of order unity. Here we take the value of massless scalar particle without affecting the result in quality.  
  ${\cal A}(\eta)$ is the factor of the power of radiation of a charged dilatonic black hole relative to the power of radiation a \sch ~black hole with the same mass. According to Fig. \ref{Fig: power Vs eta} , 
  \be {\cal A}(\eta)\in (0.038,~1],~~~  \eta \in [0, \infty). \en
  
  The second mass loss term is due to the energy loss of electrons of Schwinger effect, 
  \be 
   \frac{dM_2}{dt}=\frac{Q}{r_+}\frac{dQ}{dt},
   \en
  where rate of electron charge is determined by (\ref{volume}). The total mass loss rate is,
  \be 
  \frac{dM}{dt}=\frac{dM_1}{dt}+\frac{dM_2}{dt}.
  \label{Mevolve}
  \en
  
  Combining the evolutions of the charge and mass of a dilatonic black hole (\ref{Qevolve}) and (\ref{Mevolve}), we obtain an autonomous system,
  \be
  \frac{dQ}{dM}=f(Q,M),
  \en 
   which presents the evolution of a dilatonic black hole in the phase plane $Q-M$. 
   
   \begin{figure*}
		\centering
		\includegraphics[width=0.7\textwidth]{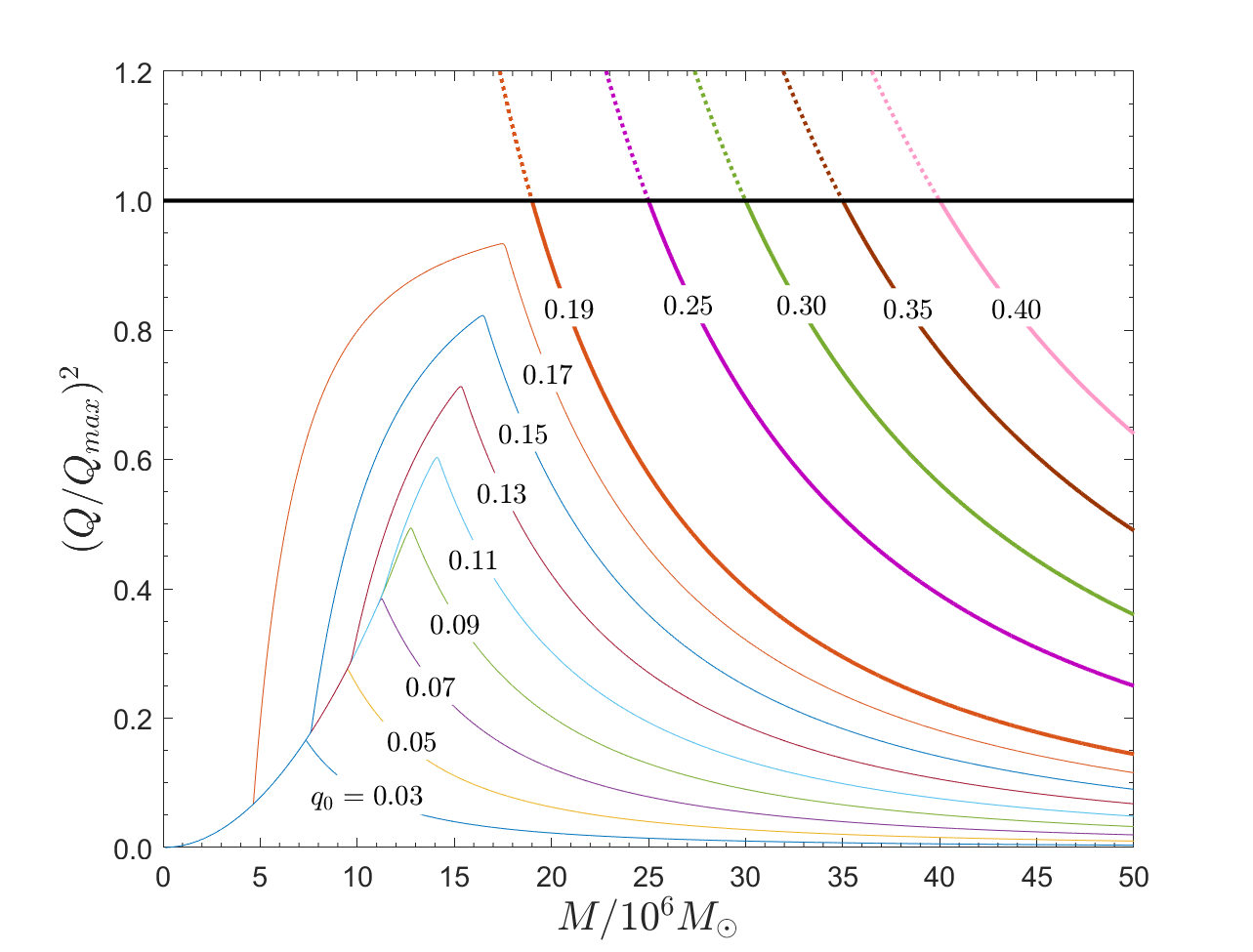}	
		\caption{Evolutions of  dilatonic black holes on the Q-M phase plane. The initial mass for the evolution of these black holes is uniformly set to $10^8M_{\odot}$, but their initial electric charges are different. Different initial charges correspond to different curves in the figure. }
		\label{Q-M}
	\end{figure*}

 We show the evolution of dilatonic black holes from different initial conditions with a unified initial mass $10^8M_{\odot}$ in Fig. \ref{Q-M}. The transition point where a naked singularity finally appears occurs at $q_0=0.17845$. For a black hole with initial charge $q_{i0}>q_0$, it inevitably evolves to be an extreme black hole. An extreme dilatonic black hole evolves to a naked singularity even if it radiates a single photon. We have demonstrated that the Hawking radiation do not cease for an extreme dilatonic black hole, while the Schwinger effects cease for an extreme dilatonic black hole. Thus the naked singularity appears spontaneously for a large dilatonic hole. This is the first deterministic evidence for the spontaneous evolution of a naked singularity from a black hole under fairly generic initial conditions.

  We would like to say some words about the evolution of a naked singularity. Any theories concerning naked singularities remain highly speculative at present. In Fig. \ref{Q-M}, we extend the theories of Hawking radiation and the Schwinger effect to the spacetime of naked singularities, as displayed by dotted curves.  According to the original demonstration of Hawking, the process of black hole evaporation depends on the existence  of event horizon. One may conjecture that the Hawking radiation, or something like that, vanishes in a spacetime without horizon. It is well-known that dynamical curved spacetime produces particles because the definition of vacuum is variable with time, for example the Friedmann-Robertson-Walker universe \cite{Birrell_Davies_1982}. Recently, a remarkable development indicates that gravitational Schwinger effects trigger particle pair productions in curved spacetimes, even for stationary cases, regardless the existence of event horizon, for which the curvature takes the status of electromagnetic field as in the Schwinger effect. As shown in the studies \cite{Wondrak:2023zdi,Ferreiro:2023jfs,Wondrak:2023hcz}, the effective temperature of gravitational Schwinger effects is the same order of Hawking temperature, and total power of gravitational Schwinger effect is about two times of Hawking effects.  
  
   The key point is that the radiation process does NOT cease for a naked singularity because of the existence curvature tensor. The concrete value of the radiation power is not critical for our fundamental discovery, that is, the inevitable violation of cosmic censorship conjecture under fairly general initial conditions.

  \section{conclusion}
	
	In this article, we study the Hawking radiations and Schwinger effects, especially the interplay of the two processes, for dilatonic black holes.  We demonstrate that for extreme dilatonic black hole, Hawking radiation behaves in a fundamentally different way compared to the case of extreme RN black hole. Under the influence of the peculiar potential at the event horizon, Hawking radiation of an extreme dilatonic black hole manifests as a truncated shrink Planck spectrum. For extreme dilatonic black holes the total power of the Hawking radiation converges to a finite value  , while the Schwinger effects vanish for large holes. Thus a naked singularity will immediately appear for an extreme dilatonic black hole through emitting Hawking photons. One should note the ‌distinctive differences‌ between the dilatonic black hole and the RN black hole: For extreme RN black hole, the power of Hawking radiation becomes zero while the Schwinger effects hold on.

  Then we further investigate whole process of the evolution of dilatonic black holes from different initial charges controlled by Hawking radiation and Schwinger effect in detail.  For a dilatonic black hole with mass $M>Q_0$, one can  throwing charge into it to force it becoming near-extremal state, and then it evolves to a naked singularity naturally. 
   Further we demonstrate that a dilatonic black hole with $10^8M_\odot$ will {\it spontaneously} evolves to a naked singularity for $q_{i0}>0.17845$. The required initial charge for emergence of naked singularity becomes smaller for a hole with larger initial masses. For the first time, we  demonstrate the inevitability of the emergence of a naked singularity under fairly general initial conditions, and under which the naked singularity arises spontaneously.

  The appearance of a naked singularity presents a dual consequence: on one hand, it places the predictive power of physics in an awkward predicament; on the other, it offers us an opportunity to directly observe singularities, potentially providing robust observational evidence for quantum gravity. Some violent phenomena in the universe, for example, gamma ray photons beyond Greisen-Zatsepin-Kuzmin (GZK) cutoff \cite{Fixsen:2009xn,Takeda:1998ps,HiRes:1993hwv}, fast radio burst (whose apparent temperature is higher than Planck temperature \cite{Keane2018}) may be attributed to radiations of naked singularities, in part.
	
	\begin{acknowledgments}
		This work is supported by the National Natural Science Foundation of China (Grants Nos.12275106 and 12235019), and the Shandong Provincial Natural Science Foundation (Grant No. ZR2024QA032).
	\end{acknowledgments}
	
	\appendix
	\label{Appendix A}
	\section{Absorption cross section of the extremal hole}
	Here, we clear some subtleties of the absorption cross section of a massless scalar field by the extreme dilatonic black hole. Following \cite{PhysRevD.14.3251}, we first consider a plane wave coming from infinity,
	\be
	\psi=e^{-i\omega t}e^{-i\omega z}.
	\en
	The current for the scalar field is given by,
	\be
	J^{\mu}=\frac{i}{2}\left(\psi^*\partial^{\mu}\psi-\psi\partial^{\mu}\psi^*\right).
	\en
	The current of the plane wave is,
	\be
	|J^z|=\omega.
	\en
	
	The number of particles absorbed by the hole per unit time is given by,
	\be
	N=\int_S \sqrt{-g}J^r dS,
	\en
	where $S$ is a surface of constant radius surrounding the black hole, and the current $J^r$ is given by
	\be
	J^r=\frac{i}{2}F(r)\left(\psi^*\partial_{r}\psi-\psi\partial_{r}\psi^*\right).
	\en
	Any solution of the wave equation in the black hole metric (\ref{Eq: metric}) can be written as,
	\be
	\psi=e^{-i\omega t}\sum_{lm}K_{lm}\frac{\psi_{\omega l}(r)}{r\sqrt{K(r)}}Y_{lm}(\vartheta,\varphi).
	\en
	Note that for scalar fields in the extremal hole metric, the asymptotic solutions of $\psi_{\omega l}(r)$ near the horizon is given by $\psi_{\omega l}\sim\mathcal{T}_{\omega l}e^{-ik_l x}$.
	Thus, we have,
	\be
	N=\sum_{lm}k_l|\mathcal{T}_{\omega l}|^2|K_{lm}|^2.
	\en
	The total absorption cross section is,
	\be\label{Eq: absorption extrem 0}
	\sigma=\frac{N}{\omega}=\sum_{lm}\frac{k_l}{\omega}|\mathcal{T}_{\omega l}|^2|K_{lm}|^2.
	\en
	Note that for the scalar field in the extremal dilatonic black hole metric, the asymptotic solution of the radial equation (\ref{Eq: the radial eq}) at the horizon and infinity are written as,
	\begin{equation}\label{Eq: absorption extrem}
	\psi_{\omega l}\sim
	\left\{
	\begin{aligned}	
	&\mathcal{T}_{\omega l}e^{-ik_l x},\;\;\;\;&\text{for}\;\;r\rightarrow r_+,\\
	&e^{-i\omega x}+\mathcal{R}_{\omega l}e^{i\omega x},\;\;\;&\text{for}\;\;r\rightarrow\infty,
	\end{aligned}
	\right.
	\end{equation}
	The conservation of the Wronskian leads to,
	\be\label{Eq: Conserv law extrem}
	|\mathcal{R}_{\omega l}|^2+\frac{k_l}{\omega}|\mathcal{T}_{\omega l}|^2=1.
	\en
	One sees that the conservation equation deviates from its ordinary form (\ref{conser}).
	
	Substituting this into Eq.(\ref{Eq: absorption extrem 0}), we obtain,
	\be
	\sigma=\sum_{lm}|K_{lm}|^2\left(1-|\mathcal{R}_{\omega l}|^2\right).
	\en
	For large $r$, one can expand the ingoing plane wave $e^{-i\omega z}$ as,
	\be
	e^{-i\omega z}=\sum_{l=0}^{\infty}\frac{i^le^{-i\omega r}}{2\omega r}\left[(4\pi)(2l+1)\right]^{1/2}Y_{l0}(\vartheta,\varphi).
	\en
	We therefore obtain,
	\be
	K_{lm}=\frac{i^l}{2\omega}\left[(4\pi)(2l+1)\right]^{1/2}\delta_{m0}.
	\en
	Finally, the absorption cross section of the extremal dilatonic black hole is given by,
	\be
	\sigma=\frac{\pi}{\omega^2}\sum_{l=0}^{\infty}(2l+1)\left(1-|\mathcal{R}_{\omega l}|^2\right),
	\en
	which is exactly the same as the non-extremal one, see Eq.(\ref{cross section}).

	\bibliography{cloud}
	
\end{document}